\documentclass[sigconf,screen]{acmart}
\acmConference[ESEM 2024]{The 18th ACM/IEEE International Symposium on Empirical Software Engineering and Measurement}{20–25 October, 2024}{Barcelona, Spain}

\AtBeginDocument{%
  \providecommand\BibTeX{{%
    \normalfont B\kern-0.5em{\scshape i\kern-0.25em b}\kern-0.8em\TeX}}}

\setcopyright{acmcopyright}
\copyrightyear{2024}
\acmYear{2024}
\acmDOI{XXXXXXX.XXXXXXX}

\usepackage{amsmath,amsfonts}
\usepackage{algpseudocode}
\usepackage{graphicx}
\usepackage[many]{tcolorbox}
\usepackage{textcomp}
\usepackage{xcolor}
\usepackage{array}
\usepackage{rotating}
\usepackage{hyperref}
\usepackage{comment}
\usepackage{xspace}
\usepackage{xcolor}
\usepackage{listings}
\usepackage{longtable}
\usepackage{paralist}
\usepackage{blindtext}
\usepackage{lscape}
\usepackage{threeparttable}
\usepackage{supertabular}

\usepackage[normalem]{ulem}
\usepackage{pifont}
\usepackage{babel,adjustbox,booktabs,multirow}
\usepackage{lscape}
\usepackage{wasysym}
\usepackage{subcaption}
\usepackage{longtable}
\usepackage{MnSymbol}

\usepackage{algorithm} 
\usepackage[algo2e,ruled,vlined,linesnumbered]{algorithm2e}

\newcommand*{\ie}{i.e.,\@\xspace}
\newcommand*{\eg}{e.g.,\@\xspace}

\newcommand{\find}[1]{
	\begin{tcolorbox}[leftrule=1mm,toprule=0mm,bottomrule=0mm,left=1pt,right=2pt,top=2pt,bottom=2pt]
		 \small #1
	\end{tcolorbox}
}

\newcommand*{\tool}{\texttt{FILLER}\@\xspace}
\newcommand*{\ST}{\texttt{SOTitle}\@\xspace}
\newcommand*{\GPT}{\texttt{GPT3.5-turbo}\@\xspace}

\newcommand*{\SO}{Stack Overflow\@\xspace}

\newcommand{\rqone}{\textbf{RQ$_1$}: \emph{
		How do Pre-trained Models influence the performance of \SO post title generation?
}}

\newcommand{\rqtwo}{\textbf{RQ$_2$}: \emph{How effective is \tool compared to state-of-the-art baselines in generating \SO post titles?}} 

\newcommand{\rqthree}{\textbf{RQ$_3$}: \emph{Which components of \tool contribute to its effectiveness?}}

\newcommand{\rqfour}{\textbf{RQ$_4$}: \emph{How are the titles generated by \tool perceived by developers compared to those obtained with \ST and \GPT?}} 

\usepackage{pifont}

\setcopyright{none}

\begin{document} \sloppy






\title[Generating SO Post Titles with PLMs, Self Improvement and Post Ranking]{Good things come in three: Generating SO Post Titles with Pre-Trained Models, Self Improvement and Post Ranking}

\author{Duc Anh Le}
\small\affiliation{%
	\institution{Hanoi University of Science and Technology}
	\city{Hanoi}
	\country{Vietnam}
}
\email{anh.ld204628@sis.hust.edu.vn}

\author{Anh M. T. Bui}
\small\affiliation{%
		\institution{Hanoi University of Science and Technology}
		\city{Hanoi}
		\country{Vietnam}
	}
\email{anhbtm@soict.hust.edu.vn}	

\author{Phuong T. Nguyen}
\small\affiliation{%
	\institution{Universit\`a degli studi dell'Aquila}
	\city{L'Aquila}
	\country{Italy}
}
\email{phuong.nguyen@univaq.it}

\author{Davide Di Ruscio}
\affiliation{%
	\institution{Universit\`a degli studi dell'Aquila}
	\city{L'Aquila}	
	\country{Italy}	
}
\email{davide.diruscio@univaq.it}

\renewcommand{\shortauthors}{ et al.}

\begin{abstract}
\textbf{Background.} \SO is a prominent Q\&A forum, supporting 
developers in seeking suitable resources on programming-related matters. Having high-quality question titles is an effective means to attract developers' attention. 
Unfortunately, this is often underestimated, leaving room for improvement. 
Research has been conducted, predominantly leveraging pre-trained models to generate 
titles 
from code snippets and problem descriptions. Yet, getting high-quality titles is still a challenging task, 
attributed to both the quality of the input data (\eg containing noise and ambiguity) and inherent constraints in sequence generation models. 


\textbf{Aims.} In this paper, we present \tool as a solution to generating \SO post titles using a fine-tuned language model with self-improvement and post ranking. \textbf{Method.} Our study focuses on enhancing pre-trained language models for generating titles for \SO posts, employing a training and subsequent fine-tuning paradigm for these models. To this end, we integrate the model's predictions into the training process, enabling it to learn from its errors, thereby lessening the effects of exposure bias. Moreover, we apply a post-ranking method to produce a variety of sample candidates, subsequently selecting the most suitable one. 
\textbf{Results.} To evaluate 
\tool, we perform experiments using 
benchmark datasets, and the empirical findings indicate that our model provides high-quality recommendations. 
Moreover, it significantly outperforms all the baselines, including \texttt{Code2Que}, \ST, \texttt{CCBERT}, \texttt{M3NSCT5}, and \GPT.
A user study also shows that \tool provides more relevant titles, with respect to \ST and \GPT. \textbf{Conclusion.} We conclude that \tool has the potential to be used in practice to support developers in generating suitable titles for their posts.

\end{abstract}




\maketitle

\section{Introduction} 
\label{sec:Introduction}

When working on software projects, developers usually seek support from different sources to complete their programming tasks~\cite{5235134}. Among others, \SO (SO) is a prominent platform when it comes to external resources for software development~\cite{10.1145/2597073.2597077,RUBEI2020106367,10.1145/2884781.2884800,DBLP:conf/se/BeyerMP021}. Essentially, developers use SO as a primary source to look for solutions to a programming task, or to find a practical way to fix errors. More importantly, they also report problems encountered during their daily tasks, asking for help and clarification from the community~\cite{DBLP:conf/se/BeyerMP021}, leveraging the so-called \emph{wisdom of the crowd.} Answers for questions posted on \SO can effectively help developers solve their problems. Nevertheless, even though there are many questions in \SO, only a small fraction of them get answered~\cite{DBLP:journals/ese/ZhuZHG22,zhou2022automatic}. This happens because the problems either need to be better described or appear unattractive to other developers~\cite{liu2022sotitle}. Among others, the title does have a role to play in the visibility of posts; existing studies \cite{DBLP:conf/cosn/CorreaS13,DBLP:journals/itc/TothV19,DBLP:conf/ecir/TrienesB19} attributed the low quality of posts to the informativeness of their titles. In fact, writing a concise and meaningful title is a challenging task, as it requires time and effort, as well as a thorough understanding of the whole post. In this respect, an automatic way to generate a title from the content of a post is highly desired.




Various studies have addressed the issue of SO post title generation~\cite{liu2022sotitle,gao2021code2que,zhang2022improving,zhang2023diverse}. 
\texttt{Code2Que}~\cite{gao2021code2que} is among the first approaches, utilizing LSTM neural networks to produce titles from source code. 
Though it achieves commendable accuracy, \texttt{Code2Que} has certain drawbacks. Specifically, it generates titles based solely on the 
snippet, overlooking the accompanying post description.
This might lead to an inability to grasp the complete context of a code snippet, causing ambiguities where the same snippet could be associated with different titles
~\cite{liu2022sotitle,zhang2023diverse}. To this end, \ST~\cite{liu2022sotitle} has been conceived to transcend such limitations. Being built on top of the Transformer structure, \ST captures long-term dependencies through a multi-head attention mechanism. More importantly, \ST is also a multitask learning tool, \ie training with posts from different programming languages. Following the same line of reasoning, various studies~\cite{gao2021code2que,zhang2022improving,zhang2023diverse} also deal with multitask learning, exploiting various pre-trained language models, 
and obtaining an encouraging accuracy. 


However, as we 
show later on in Section~\ref{sec:Challenges}, there are two main difficulties 
in post title recommendation: \emph{(i)} The potential for exposure bias, which mainly emerges from differences in how the model is trained compared to how it operates during inference; and \emph{(ii)} The inherent randomness in sequence generation models, leading to significant fluctuations in the quality of generated titles. Altogether, this poses challenges in providing relevant recommendations. 

\begin{figure*}[t!]
	\centering
	\includegraphics[width=0.78\textwidth]{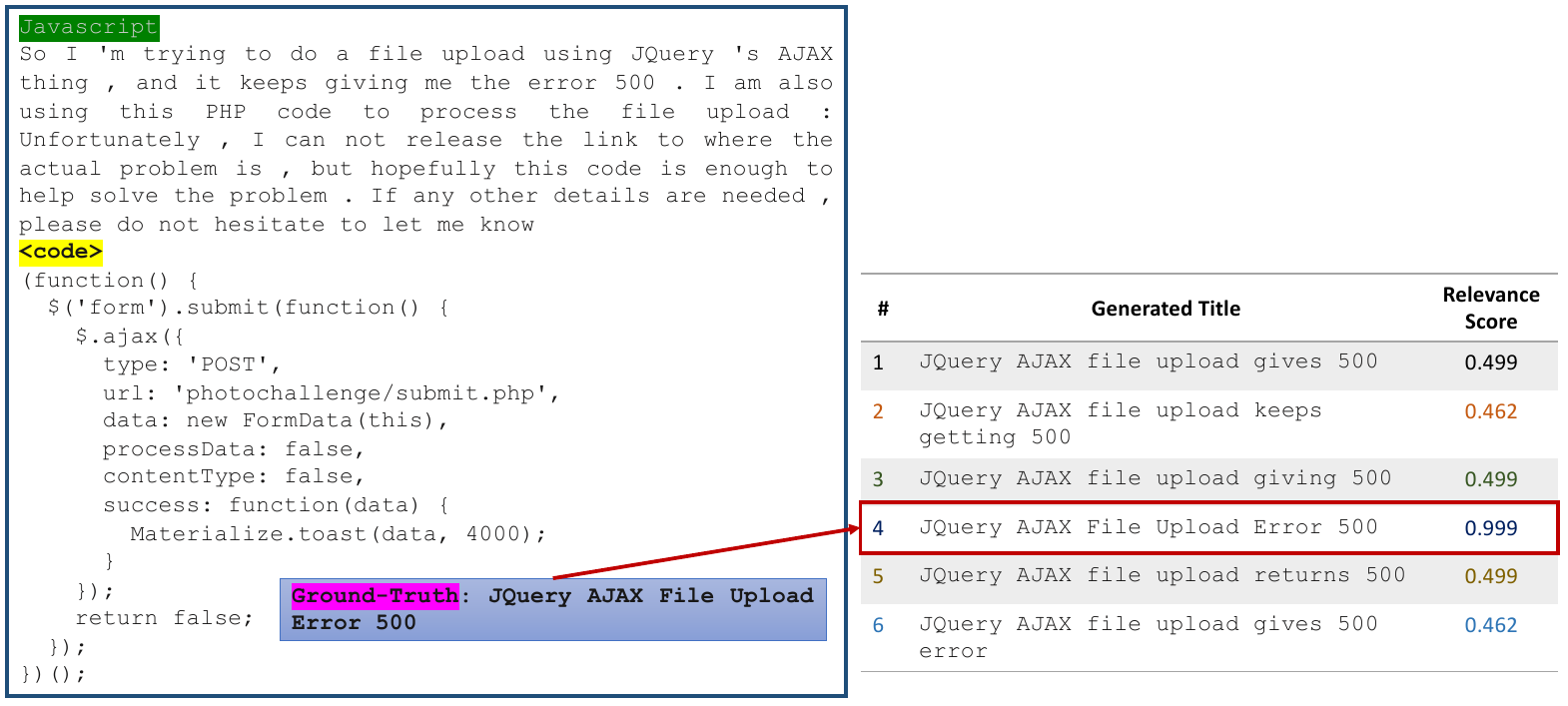}
	\caption{Titles generated by CodeT5: The firstly ranked title is not the most relevant one.}
	\label{fig:ranking-example}
\end{figure*}

This paper presents a practical approach to generating titles for \SO posts. 
We conceptualize a tool named \tool to overcome the aforementioned obstacles in SO post title generation, leveraging 
\textbf{FI}ne-tune \textbf{L}anguage mode\textbf{L} with self improv\textbf{E}ment and post \textbf{R}anking. %
An empirical evaluation using real-world datasets that cover posts on 4 different programming languages demonstrated that \tool is highly effective in providing recommendations. We compared \tool with 5 state-of-the-art approaches for post title recommendation, including 
\texttt{Code2Que}~\cite{gao2021code2que},
\ST~\cite{liu2022sotitle},  \texttt{CCBERT}~\cite{zhang2022improving}, \texttt{M3NSCT5}~\cite{zhang2023diverse}, and \GPT~\cite{openai2023gpt35turbo}. More importantly, we evaluate \tool against \ST and \GPT by means of a user study. The results of the experiment demonstrate that \tool obtains a promising 
performance, outperforming all the competitors. 
In this respect, our paper makes the following contributions:


	\smallskip
	\noindent
	$\triangleright$ \textit{Investigation.} To the best of our knowledge, this work is the first one ever that has examined the applicability of fine-tuning language model with self improvement and post ranking in generating SO titles. 

    \smallskip
    \noindent
    $\triangleright$ \textit{Solution.} We develop \tool, a novel tool for automatically generating titles for \SO posts using their content with code and text as input, attempting to create concise and informative titles for the posts under consideration. 
    
    \smallskip
    \noindent
    $\triangleright$ \textit{Evaluation.} An empirical evaluation together with a user study has been conducted using real-world datasets to study the performance of our proposed approach, and compare it with five state-of-the-art baselines, \ie \texttt{Code2Que}~\cite{gao2021code2que}, \ST~\cite{liu2022sotitle}, \texttt{CCBERT}~\cite{zhang2022improving} and \texttt{M3NSCT5}~\cite{zhang2023diverse}, 
    and \GPT~\cite{openai2023gpt35turbo}. 
    
    \smallskip
    \noindent
    $\triangleright$ \textit{Open Science.} To facilitate future research, we publish a replication package including the dataset and source code of \tool~\cite{FILLERartifacts}.

 \smallskip
\noindent
\textbf{Structure.} 
Section~\ref{sec:RelatedWork} reviews the related work and introduces the research motivation. We describe in detail the proposed approach in Section~\ref{sec:Solution}. Afterward, Section~\ref{sec:settings} presents the materials and methods used to conduct an empirical evaluation on \tool. 
Section~\ref{sec:Results} reports and analyzes the experimental results, it also highlights 
the threats to validity of our findings. 
Finally, Section~\ref{sec:Conclusion} sketches future work and concludes the paper.

\section{BACKGROUND AND MOTIVATION} 
\label{sec:RelatedWork}

This section reviews state-of-the-art research dealing with the problem of generating SO posts. Based on this, we also identify the existing challenges and develop proposals to overcome them.

\subsection{Related Work}


Mondal et al.~\cite{mondal2021early} highlighted that an increasing number of open questions on SO remain unanswered, partially attributed to the suboptimal quality of the question expression. Numerous research endeavors have been dedicated to addressing this issue, with a specific focus on crafting high-quality question titles~\cite{gao2021code2que,zhang2022improving,liu2023automated,zhang2023diverse,liu2022sotitle,zhou2023qtc4so}.
These investigations predominantly revolved around generating question titles by analyzing the content of SO posts, including problem descriptions and code snippets.

Initial studies in this domain depended solely on code snippets for generating question titles, aligning closely with the code summarization task~\cite{tang2022ast,cheng2022keyword,shi2022natural,zhou2022automatic}. 
Gao et al.~\cite{gao2021code2que} introduced \texttt{Code2Que} to generate SO titles. 
This model employs bidirectional LSTMs, coupled with an attention mechanism, to encode the sequence of code tokens into hidden states and subsequently decode them into concise question titles in natural language. 
This method also incorporated two conventional Natural Language Processing (NLP) techniques. It implemented a coverage mechanism~\cite{tu2016modeling} to mitigate word repetition issues and a copy mechanism~\cite{gu2016incorporating} to address out-of-vocabulary challenges.
Though \texttt{Code2Que} exhibits promising performance, LSTMs may struggle to handle long-range dependencies, leading to challenges in capturing the semantic meaning of the entire source code.
The most recent research efforts seek to harness the extensive dataset for pre-training language models and then fine-tuning for a specific task, resulting in state-of-the-art achievements, particularly in code summarization~\cite{ahmad2021unified,guo2022unixcoder} and SO post title generation~\cite{zhang2023diverse}.
Ahmad et al.~\cite{ahmad2021unified} introduced a Transformer model, PLBART, to tasks related to programming languages such as code summarization, code generation and code translation.
Subsequent studies focus on introducing multilingual representation models tailored for programming languages, as exemplified by UnixCoder~\cite{guo2022unixcoder}, CodeT5~\cite{wang2021codet5}.
Zhang et al.~\cite{zhang2023diverse} introduced \texttt{M3NSCT5}, fine-tuning the CodeT5 model to translate code sequences into relevant titles for SO posts. 
However, the lack of context around code snippets can lead to the same code snippets being associated with different titles, introducing ambiguity. 
In addressing this, \texttt{M3NSCT5} emphasizes diversifying generated texts using Nucleus Sampling strategy~\cite{holtzman2019curious} and Maximal Marginal Ranking~\cite{yang2021maximizing}, providing users with a selection of titles for a given code snippet.
While our approach shares the aspect of producing multiple candidates during the inference stage, our primary goal differs–we aim to identify the most relevant title aligned with the input question description and code snippets.

Recent studies in 
code summarization also proved that the inclusion of additional information, such as API documentation~\cite{hu2018summarizing}, code comments~\cite{wei2020retrieve}, alongside the source code, has demonstrated improved performance compared to relying solely on the source code.
Pre-trained models with the ability of multi-modal input modelling are typically adopted in these studies.
Zhang et al.~\cite{zhang2022improving} introduced \texttt{CCBERT}, which leverages the pre-trained CodeBERT model~\cite{feng2020codebert} to parse bi-modal contents (i.e., problem description and code snippets), incorporated with the copy mechanism~\cite{gu2016incorporating} to handle rare tokens.
Liu et al.~\cite{liu2022sotitle} also incorporate multi-modal modeling in their approach, presenting \ST for generating SO post titles.

\ST was constructed upon the pre-trained T5 model~\cite{raffel2020exploring}, aiming to establish a unified model for concurrent training across multiple programming languages. Subsequent research endeavors have adopted this input representation approach for Stack Overflow post title generation~\cite{zhang2023diverse,liu2023automated,zhou2023qtc4so}.
Instead of creating new titles for SO posts, Liu et al.~\cite{liu2023automated} focused on reformulating question titles through the analysis of modification logs on the Stack Overflow platform.
The proposed \texttt{QETRA} model also employs the pre-trained T5 as the backbone model to learn the title reformulation patterns from the extracted title modification logs.
In contrast to previous research, \texttt{QETRA} employs both the question body (comprising code snippets and problem description) and the question titles provided by users to suggest candidate reformulated titles.
Another task closely related to the generation of Stack Overflow post titles is the completion of question titles.
Zhou et al.~\cite{zhou2022automatic} introduced \texttt{QTC4SO}, an innovative approach for generating Stack Overflow post titles from bi-modal post contents and incomplete titles. Utilizing the pre-trained T5 model, \texttt{QTC4SO} learns potential title completion patterns and incorporates a multi-task learning strategy to leverage meaningful information from various programming languages.

\subsection{Challenges} \label{sec:Challenges}

When utilizing pre-trained models (PTMs) for natural language generation tasks in general and specifically for creating SO post titles, two crucial issues need attention to enhance the quality of the generated text as follows.  %
The first issue (I1) concerns the potential for exposure bias, which mainly stems from differences in how the model is trained compared to how it operates during inference. During training, the model is designed to predict the next word in a sequence using the correct prior words from the training data, ensuring that it always operates within the correct contextual framework up to that point. However, during inference, the model generates words sequentially, relying on its own prior predictions to inform subsequent ones. As a result, any errors in early predictions can be compounded, leading to outputs that lack coherence. Enhancing the training dataset with examples that mirror the conditions experienced during inference can aid the model in learning to manage its own mistakes.

\begin{figure*}[t!]
	\centering
	\includegraphics[width=0.80\textwidth]{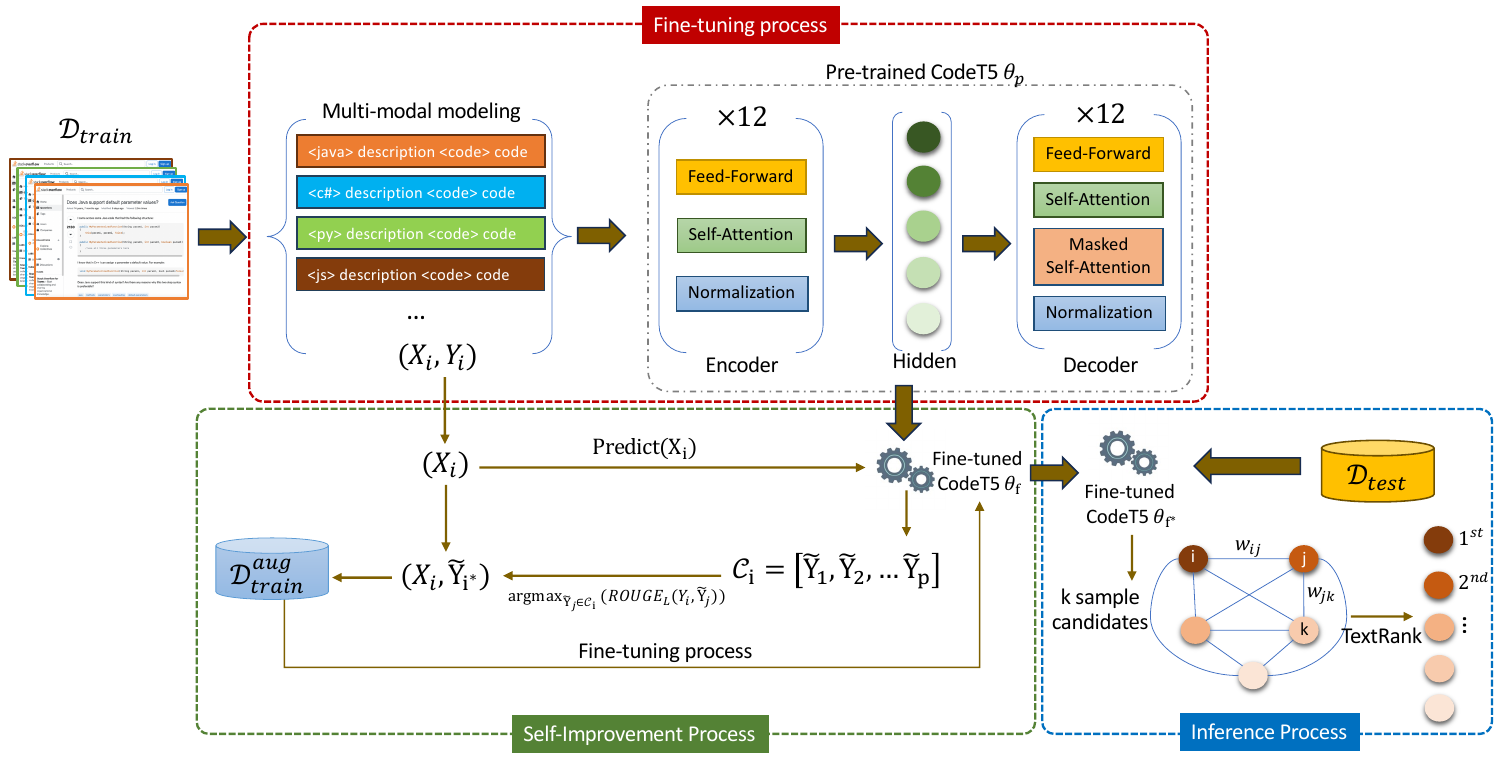}
	\caption{The architecture of \tool.}
	\label{fig:Architecture}
	\vspace{-.2cm}
\end{figure*}

The second issue \emph{(I2)} pertains to the inherent randomness in sequence generation models, leading to significant fluctuations in the quality of generated titles. 
Figure~\ref{fig:ranking-example} depicts a motivating example, where CodeT5 was used to generate a title for a specific post,\footnote{The post is found in the following link: \url{https://stackoverflow.com/questions/30010684/jquery-ajax-file-upload-error-500}} which is about asking for a solution to a JavaScript snippet. 
The post already had an existing title: \emph{``JQuery AJAX File Upload Error 500,''} serving as the ground-truth data. We generated 30 title candidates using 
CodeT5 
and computed their relevance scores using the TextRank algorithm~\cite{DBLP:conf/emnlp/MihalceaT04}. Remarkably, the results of our experiment showed that the first title produced by CodeT5 was not the most relevant one, with a relevance score of only 0.499. However, the 4th sample on the list perfectly matches the ground-truth data, earning a TextRank relevance score of 0.999. This example highlights the importance of generating multiple 
candidates to increase the likelihood of achieving high-quality titles. 
This method has been successfully implemented in several studies, \eg those conducted to improve the performance of fault-aware neural networks~\cite{inala2022fault} and neural language to code translators \cite{shi2022natural}.

To overcome the aforementioned obstacles in 
post title generation, we develop \tool on top of 
fine-tune language model with self improvement and post ranking. 
To tackle 
\textit{I1}, we 
augment the training dataset by integrating the model's own predictions. Employing predictions as inputs during training may reduce the discrepancy between training and inference, 
enhancing the model's adaptability to real-world data.
For 
\textit{I2}, we employ a post-ranking method that produces a variety of 
candidates, subsequently selecting the most suitable one,
thus increasing the likelihood of generating more relevant titles. 

\section{PROPOSED SOLUTION}
\label{sec:Solution}


This section describes our proposed approach to \SO post title generation--\tool--whose overall workflow is depicted in Figure~\ref{fig:Architecture}.
There are three main stages as follows: \emph{fine-tuning, self-improvement} and \emph{inference}. During the fine-tuning stage, each SO post, comprising a question description and code snippet, is combined to form a bi-modal input.
Our approach adopts a multi-task learning strategy, similar to previous studies~\cite{liu2022sotitle,zhang2023diverse,liu2023automated}, training the model simultaneously on various programming languages. We utilize CodeT5~\cite{wang2021codet5} 
as the backbone model and proceed with fine-tuning it on a \SO post dataset.

In addition to the conventional  pre-training and \emph{fine-tuning} paradigm, our goal is to boost the fine-tuned model's performance by augmenting the training dataset. This enhanced dataset will be employed in a further fine-tuning stage. 
This process is referred to as the {\it self-improvement} technique~\cite{to-etal-2023-better}.
In the final \textit{inference} stage, the model generates multiple candidate samples rather than producing just a single sample. Our proposed post-ranking method is designed to identify and select the highest quality title from these.
%

\subsection{Fine-tuning PTM with multi-modal inputs} \label{sec:plms}
Fine-tuning using a PTM for downstream tasks is a prevalent paradigm, extensively employed in numerous SE tasks~\cite{liu2023refining,chow2023beware,le2023invalidator}.
\tool also follows this paradigm, utilizing the pre-trained CodeT5~\cite{wang2021codet5} model as the backbone, then further updating its trainable parameters using a task-specific dataset consisting of pairs of post title and body.

Getting inspiration from previous research~\cite{zhang2022improving,liu2022sotitle,zhang2023diverse}, we create a bi-modal input for each SO post by merging the code snippet with the question description. 
This combined representation is then tokenized using the Byte-Pair Encoding (BPE) method~\cite{sennrich-etal-2016-neural}, which was integrated into CodeT5. 
This tokenizer is specifically designed to tackle the Out-of-Vocabulary (OoV) issues efficiently. We also explore a multi-task learning setting that enables training a shared model on multiple programming languages (PLs) simultaneously. This approach allows for the reuse of model weights across various PLs, thereby reducing computational costs and enhancing the model's generalization capability~\cite{liu2020multi}. 
We prepend a prefix to the input sequence to inform the model about the specific language it is dealing with. 
Given a post, denoted as $\mathcal{X}$ which includes a sequence of code ($X_{code}$) and a sequence of question description ($X_{desc}$), the multi-modal input of $\mathcal{X}$ is formulated as shown in Equation~\ref{eq:input}.
\begin{equation}
    \label{eq:input}
    \mathcal{X} \,=\, <prefix> \oplus \,X_{desc}\, \oplus \,<code>\, \oplus\, X_{code}
\end{equation}
where $<prefix>$ denotes the programming language--specific prefix. For example, the prefix \texttt{JS} indicates the JavaScript programming language.
The special separator $<code>$ is used to distinguish between $X_{desc}$ and $X_{code}$.

We fine-tune 
CodeT5~\cite{wang2021codet5} 
using the formatted SO posts. CodeT5 is built on top of the Encoder-Decoder architecture~\cite{vaswani2017attention}, with 12 blocks of feed-forward neural networks, self attention, and normalization by each size (\ie $\times 12$).
 The goal of this stage is to identify a set of parameters $\theta$ that minimizes the negative log-likelihood of the target title tokens $Y=\{y_t\}$  when conditioned on the corresponding input sequence $\mathcal{X}$ from the training dataset, given below. 
\begin{equation}
    \label{eq:finetune}
    \mathcal{L}_{\theta} = - \sum\limits_{t=1}\limits^{|Y|} \log P_\theta(y_t|y_{<t};\mathcal{X})
\end{equation}
The optimal parameter $\theta_f$ is determined in Equation~\ref{eq:theta}.
\begin{equation}
\label{eq:theta}
\theta_f = \arg\min\limits_{\theta}\frac{1}{N}\sum_{\ell=1}^N\mathcal{L}^{\ell}_\theta
\end{equation}
where $\mathcal{L}^{\ell}$ denotes the loss function for the corresponding programming language $\ell$, $N$ is the number of the programming languages. In this study, we examine \SO posts from four programming languages including \texttt{Java}, \texttt{C\#}, \texttt{Python} and \texttt{JavaScript}.
During the inference phase, we use the fine-tuned model in conjunction with the auto-regressive decoding method to get the predicted title $\tilde{Y}$ token by token.

\subsection{Self Improvement}
\label{sec:self}

Pre-trained models have shown considerable versatility in handling diverse tasks, greatly diminishing the need for extensive engineering by enabling multi-modal modeling~\cite{wang2023large}, 
However, these models typically encounter a potential exposure bias due to the misalignment between the training conditions and those during inference~\cite{wang2020exposure,guerreiro2023hallucinations,to-etal-2023-better}.
To alleviate such a bias, incorporating the model's own predictions during training can be beneficial~\cite{wang2020exposure}. This strategy allows the model to better manage its errors, leading to enhanced performance.
Inspired by this approach, we enhance our training dataset by incorporating predictions from the initially fine-tuned model. This augmented dataset is then employed to further refine the model through an additional fine-tuning stage.

The self-improvement strategy is illustrated in Algorithm~\ref{alg:si}. 
Its input includes the initial training dataset $\mathcal{D}_t$, the number of generated candidates $k$ and the initially fine-tuned CodeT5 model ($\mathcal{M}_{\theta_f}$) which is characterized by the set of parameters $\theta_f$.
This procedure yields an enhanced version of the training dataset, denoted as $\mathcal{D}^{aug}_t$, which is then employed to fine-tune the model further.
For each pair consisting of a post and its 
ground truth title, $k$ title candidates are generated using the initially fine-tuned CodeT5 model. The candidate that achieves the highest ROUGE--L score 
when compared to the ground truth title is selected for inclusion in the augmented training dataset (Lines 2--9).

\begin{algorithm2e}[t]
\caption{Pseudo code of the Self--Improvement Process}
\label{alg:si}
\DontPrintSemicolon
\SetAlgoLined
\SetKwInOut{Input}{Input} 
\SetKwInOut{Output}{Output}
\Input{
    \begin{itemize}
        \item The fine--tuned model $\mathcal{M}_{\theta_f}$
        \item The original training dataset $\mathcal{D}_{t}$
        \item $k$ that denotes the number of generated candidates
    \end{itemize}
}
\Output{
\begin{itemize}
    \item The augmented training dataset $\mathcal{D}^{aug}_{t}$
    \item The new fine-tuned model $\mathcal{M}_{\theta_{f^*}}$
\end{itemize}
}

$\mathcal{D}^{aug}_{t} \leftarrow \emptyset$

\ForEach {$(\mathcal{X}_i,Y_i) \in \mathcal{D}_t$}
{
    $\mathcal{C}_i \leftarrow \emptyset$
    
	\ForEach {$j \in [1..k]$}
	{
		$\tilde{Y}_j \leftarrow \mathcal{M}_{\theta_f}\text{.predict}(\mathcal{X}_i)$
		
		$\mathcal{C}_i \leftarrow \mathcal{C}_i \bigcup \tilde{Y}_j$
	}

    $\tilde{Y}^*_i \leftarrow \text{arg}\max\limits_{\tilde{Y}_j \in \mathcal{C}_i} ROUGE_L(Y_i,\tilde{Y}_j)$
	
	$\mathcal{D}^{aug}_t \leftarrow \mathcal{D}^{aug}_t \bigcup (\mathcal{X}_i,\tilde{Y}^*_i)$
}

$\mathcal{M}_{\theta_{f^*}} \leftarrow \text{Fine-Tuning}( \mathcal{M}_{\theta_f},\mathcal{D}^{aug}_t)$

\Return{$\mathcal{D}^{aug}_t$ and $\mathcal{M}_{\theta_{f^*}}$}
\end{algorithm2e}

\subsection{Post Ranking}
\label{sec:ranking}
Blending ground truth with model predictions during the training phase is essential for better alignment with the inference process. 
Yet, in the text generation process, sequence generation models often utilize decoding strategies like beam search, greedy search, and nucleus sampling to select the best sequence of tokens~\cite{holtzman2019curious,zhang2023diverse}. 
These methods’ inherent randomness can lead to significant variability in the quality of generated text. 
A widely used solution to this problem involves generating a range of sample candidates and then selecting the highest quality title among them~\cite{shi2022natural,inala2022fault}.

In this research, our goal is to generate high-quality titles for \SO posts, with a particular emphasis on addressing the issue of high variability in generation quality by implementing a post-ranking method during the inference phase. For each \SO post input $\mathcal{X}$, the model generates a set of $\mathbb{K}$ title candidates, denoted as $\mathcal{T} = \{T_1, T_2, \dots, T_{\mathbb{K}}\}$. To assess the relevance of these generated titles, we apply the TextRank algorithm, an unsupervised, graph-based ranking technique. In this approach, we construct a graph $\mathcal{G}=\{\mathcal{T},\mathcal{E}\}$, where each title candidate $T_i \in \mathcal{T}$ is a node in the graph $\mathcal{G}$. 
We initialize an edge between every pair of nodes, with the weight of each edge determined by the cosine similarity between the two corresponding title candidates.
To calculate the cosine similarity between two title candidates, we utilize the Term Frequency and Inverse Document Frequency (TF--IDF) method. We construct a token vocabulary $\mathcal{V}$ from the set of titles $\mathcal{T}$. Each title $T_i$ is represented by a vector $v_{T_i} = {w_1^i, w_2^i,\dots,w_m^i}$, where $m$ is the size of the vocabulary, and $w_j^i$ is calculated as follows:
\begin{equation}
    \label{eq:tf-idf}
    w_j^i = tf_j^i \times idf_j = (\log{f_j^i}+1) \times (\log{\frac{\mathbb{K}}{\mathbb{K}_i}} + 1)
\end{equation}
with $f_j^i$ being the frequency of the $j^{th}$ token in title $T_i$, $\mathbb{K}_i$ as the number of titles that contain this token, and $\mathbb{K}$ being the total number of title candidates. 
The lexical similarity between the TF--IDF vectors of $T_i$ and $T_j$ is computed, as detailed in Equation~\ref{eq:ranking}.
\begin{equation}
    \label{eq:ranking}
    Sim(T_i,T_j) = \ell_{ij} =\frac{\overrightarrow{v_{T_i}} \cdot \overrightarrow{v_{T_j}}}{{\lVert\overrightarrow{v_{T_i}}\rVert\times\lVert\overrightarrow{v_{T_j}}\rVert}} 
\end{equation}
where $v_{T_i}$ and $v_{T_j}$ denote the TF--IDF vectors of $T_i, T_j$, respectively.
The TextRank algorithm assigns a relevance score to each node, which is iteratively refined until it converges, as in Equation~\ref{eq:textrank}.
\begin{equation}
    \label{eq:textrank}
    \mathcal{S}_{i} = (1-a) + a \sum\limits_{j \neq i}  \frac{\ell_{ij}}{\sum\limits_{k \neq j}\ell_{jk}}\mathcal{S}_j
\end{equation}
where $a$ represents a damping factor used in the TextRank algorithm, typically set to $0.23$. $\mathcal{S}_i$ and $\mathcal{S}_j$ denote the relevance scores of the $i^{th}$ and $j^{th}$ nodes, respectively.
The title that achieves the maximum relevance score is then selected as the best one.

\section{EMPIRICAL EVALUATION}  
\label{sec:settings}

In this section, we present the empirical evaluation conducted to study the performance of our proposed approach.

\subsection{Research Questions}
\label{sec:ResearchQuestions}

A series of experiments was performed to answer the following research questions:

\vspace{.2cm}
\noindent \rqone~This research question seeks to determine the efficacy of Pre-trained Models (PTMs) in generating titles for SO posts. 
We evaluate five distinct PTMs, namely BART~\cite{lewis2020bart}, T5~\cite{raffel2020exploring}, CodeT5~\cite{wang2021codet5}, CodeBERT~\cite{feng2020codebert} and UnixCoder~\cite{guo2022unixcoder}.
The first two models are designed for natural language processing, whereas the latter three are specialized in understanding and representing programming languages.
We applied the fine-tuning process to such pre-trained models utilizing the benchmark datasets described in Section~\ref{sec:dataset}, and by following the 
settings outlined by Liu et al.~\cite{liu2022sotitle}.

\vspace{.2cm}
\noindent \rqtwo~In this question, we compare the performance of our proposed approach to the baseline works. We choose four recent state-of-the-art models for \SO post title generation including \texttt{Code2Que}~\cite{gao2021code2que}, \texttt{SOTitle}~\cite{liu2022sotitle}, \texttt{CCBERT}~\cite{zhang2022improving} and \texttt{M3NSCT5}~\cite{zhang2023diverse}, which have obtained a promising recommendation performance. 

    
    \smallskip
    \noindent
    $\triangleright$ \texttt{Code2Que}: 
    Gao et al.~\cite{gao2021code2que} proposed the \texttt{Code2Que} model to generate Stack Overflow titles from code snippets. This model is underpinned by an LSTM encoder-decoder architecture, augmented with the integration of copy and coverage mechanisms. 
    
	\smallskip
	\noindent
	$\triangleright$ \texttt{SOTitle}: 
    Liu et al.~\cite{liu2022sotitle} fine-tuned the pre-trained T5 model on their collected \SO dataset. 
    They employed a multi-task learning architecture in which the description and code snippet of each SO post were concatenated together for each programming language but followed a simultaneous training process. 
    
    \smallskip
    \noindent
    $\triangleright$ \texttt{CCBERT}: 
    The model was proposed by Zhang et al.~\cite{zhang2022improving}, employing CodeBERT to encode both the code snippets and question descriptions. It is further enhanced by integrating a copy attention layer to refine the generated output's quality.
    
    \smallskip
    \noindent
    $\triangleright$ \texttt{M3NSCT5}: 
    This model was developed to generate multiple post titles from the given code snippets~\cite{zhang2023diverse}. The pre-trained CodeT5 model was utilized as the backbone, combined with the maximal marginal ranking strategy to select the most relevant and diverse titles from multiple generated samples.
    
    \smallskip
    \noindent
    $\triangleright$ \GPT: To the best of our knowledge, 
    no-one has ever applied a large language model (LLM) to post title generation. In this work, we examine how well \GPT~\cite{openai2023gpt35turbo}--an LLM--can generate suitable titles. 
    To this end, we used a paid account from ChatGPT, and ran the experiments with its API, querying the service using tokens. 
    

It is worth mentioning that all the baselines mentioned earlier--except \texttt{Code2Que}--are predominantly fine-tuned pre-trained language models for generating post titles. Notably, \texttt{SOTitle} and \texttt{CCBERT} are bi-modal models, whereas \texttt{Code2Que} and \texttt{M3NSCT5} rely solely on code snippets. 
For a fair comparison with \tool, 
we fine-tuned these models using 
snippets and question descriptions, following the methodology 
by Liu et al~\cite{liu2022sotitle}.

\vspace{.2cm}
\noindent \rqthree~In our work, we implemented two strategies to enhance the \SO post title generation model: a Self-Improvement approach during training and a Post-Ranking method for selecting the most pertinent title for a given question body. By means of an ablation study, this research question assesses how each of these components contributes to the overall efficacy of the proposed model.  

\vspace{.2cm}
\noindent \rqfour~
To address the limitations of automatic 
metrics such as ROUGE~\cite{10.1145/3597503.3639174},
which only assess the lexical overlap between the reference and generated titles, we ran a human study to evaluate the performance of \tool compared to 
\ST and \GPT. We chose them 
due to the following reasons: \emph{(i)} It is impractical to compare \tool with all the baselines, as the number of samples to be examined 
is large; thus \emph{(ii)} \ST was selected as it is among the well-established approaches, with source code implementation, allowing us to tailor the experiments; 
and \emph{(iii)} \GPT is an LLM, having the potential to recommend relevant post titles.

Following existing work~\cite{zhang2022improving}, this research primarily assesses two key aspects of the generated titles: \emph{(i)} \emph{Readability} and \emph{(ii)} \emph{Relevance}, with each aspect being rated on a scale from 1 to 4 
(see Table~\ref{tab:human-metrics}). 
The former evaluates the grammatical correctness and fluency of a generated title, and this is done just by reading the title. Meanwhile, the latter measures how relevant the generated title is with respect to the actual content of the original post description as well as the ground-truth title.

\begin{table}[t!]
    \centering
    \footnotesize
    \caption{Designed human evaluation criteria~\cite{zhang2022improving}.}
    \begin{tabular}{|l|p{5.7cm}|c|}
    \hline 
         \textbf{Criteria}  & \textbf{Description} & \textbf{Score} \\ \hline
        	& \multicolumn{2}{l|}{\textit{We evaluated if the generated title:}} \\ \cline{2-3}
           	{\multirow{5}{*}{\rotatebox[origin=c]{90}{\textbf{Readability}}}}  &  - contains numerous grammatical errors, making it difficult to read and understand & 1 \\ \cline{2-3}
            &-~contains a few grammatical errors but remains readable and understandable & 2 \\ \cline{2-3}
            &-~is easy to follow and read with minimal grammatical errors & 3 \\ \cline{2-3}
            &-~is exceptionally well-written, clear, and appealing in terms of grammatical accuracy & 4 \\ \hline
			{\multirow{5}{*}{\rotatebox[origin=c]{90}{\textbf{Relevance}}}} & \multicolumn{2}{l|}{\textit{The generated title in relation with the original post:}} \\ \cline{2-3}
         	 &-~It misses completely the essence of the post's content & 1 \\ \cline{2-3}
            &-~It partially reflects the main points of the post & 2 \\ \cline{2-3}
            &-~It aligns well with the key points of the post & 3 \\ \cline{2-3}
            &-~It summaries perfectly the content of the post & 4 \\  \hline
    \end{tabular}
    \label{tab:human-metrics}
    \vspace{-.2cm}
\end{table}

We randomly selected 200 samples from the testing dataset, obtaining 600 titles generated by the three models. For the evaluation, we involved three master students in Computer Science--who are not co-authors of this study--in manually assessing the titles. The students are familiar with the \SO platform, and have experience with programming in different languages. Aiming for a fair comparison, we followed existing guidelines in conducting a user study~\cite{DBLP:journals/ese/RoccoRSNR21}. In particular, we anonymized the origin of the titles, so as to conceal the actual tool that generates them. Each student was assigned 200 questions, and they evaluated both \emph{Readability} and \emph{Relevance} by reading and comparing the generated titles with the ground-truth posts. Once the students finished with the process, two senior developers--who are co-authors of this paper--were then asked to validate the results, re-evaluate, 
and discuss with the students 
to resolve any possible inconsistencies, finally reaching a consensus. 
It is worth noting that all the discussion was done also on anonymized post titles, aiming to avoid any bias or prejudice against any specific tools.

\subsection{Dataset}\label{sec:dataset}
We use a benchmark dataset from a previous study~\cite{liu2022sotitle}, comprising \SO posts in four different languages: Java, C\#, Python, and JavaScript (JS), which are among the most popular programming languages. Each post includes a brief title, a descriptive question, and a code snippet. The dataset encompasses 284,298 posts across these four languages, divided into training, validation, and testing subsets. For each language, there are 60,000 and 5,000 posts for training and testing, respectively. Detailed statistics of the benchmark dataset are shown in Table~\ref{tab:dataset}.

\begin{table}[h!]
	\small
    \centering
    \caption{Statistics of the benchmark dataset.}
    \begin{tabular}{lccc}
    \hline
         \textbf{Language} &  \textbf{Training}  & \textbf{Validation}& \textbf{Testing} \\
         \hline
          Java & 60,000 & 3,959 & 5,000 \\
          C\#  & 60,000 & 6,817 & 5,000 \\
          Python & 60,000 & 7,742 & 5,000 \\
          JavaScript & 60,000 & 5,780 & 5000 \\ \hline
          \textbf{Total} & 240,000 & 24,298 & 20,000 \\ \hline
    \end{tabular}
    \label{tab:dataset}
    \vspace{-.2cm}
\end{table}

  
\subsection{Evaluation Metrics}\label{sec:metric} 
To evaluate the accuracy of the generated Stack Overflow post titles, we utilize ROUGE metrics~\cite{lin2004rouge}. Specifically, we calculate ROUGE-1 (R-1), ROUGE-2 (R-2), and ROUGE-L (R-L), which have been extensively used in prior studies for tasks involving text generation and summarization~\cite{liu2022sotitle,gao2021code2que,zhang2022improving,zhang2023diverse}. 
ROUGE--1 and ROUGE--2 rely on 1--grams and 2--grams, respectively, whereas ROUGE--L focuses on the longest continuous sequence.
The recall, precision, and F1-Score for the ROUGE-$k$ metrics are defined below. 

\begin{equation} \nonumber
    \label{eq:recall}
    \text{Recall}_{rouge-k} = \frac{\#\text{overlapped\_}k\text{\_grams}}{\#k\text{\_grams} \in \text{gold summary}}
\end{equation}

\begin{equation} \nonumber
    \label{eq:precision}
    \text{Precision}_{rouge-k} = \frac{\#\text{overlapped\_}k\text{\_grams}}{\#k\text{\_grams} \in \text{generated summary}}
\end{equation}

\begin{equation} \nonumber
    \label{eq:f1}
    \text{F1-Score}_{rouge-k} = 2 \times \frac{\text{Recall}_{rouge-k} \times \text{Precision}_{rouge-k}}{\text{Recall}_{rouge-k} + \text{Precision}_{rouge-k}}
\end{equation}
where the {\it gold summary} refers to the original title of the SO post while the {\it generated summary} denotes the title generated by the underlying model.
The {\it overlapped\_$k$\_grams} indicates the $k$\_grams which appears in both the gold summary and the generated one. 
The recall metric calculates the proportion of $k$-grams in the reference summary that are included in the generated summary, whereas precision determines the proportion of $k$-grams in the generated summary that are present in the gold summary. The F1-Score represents a balance between recall and precision.
This study adopts the F1-Score as the primary evaluation metric, in line with prior research~\cite{zhang2022improving,zhang2023diverse}.
Additionally, we employ the recall metric to compare how different models perform in aligning the generated summaries with the reference ones in terms of similarity~\cite{liu2022sotitle}.

\begin{table*}[htp]
    \centering
    \small
    \caption{Performance of Pre-trained Models (PTMs) including BART, T5, CodeBERT, CodeT5, and UnixCoder in title generation for \SO Posts.}
    \begin{tabular}{l|ccc|ccc|ccc|ccc|ccc}
        \hline
        \multirow{2}{*}{\textbf{Model}} & \multicolumn{3}{c|}{\textbf{Java}} & \multicolumn{3}{c|}{\textbf{C\#}} & \multicolumn{3}{c|}{\textbf{Python}} & \multicolumn{3}{c|}{\textbf{JavaScript}} & \multicolumn{3}{c}{\textbf{Average}}\\
        \cline{2-16} & R--1& R--2& R--L& R--1& R--2& R--L& R--1& R--2& R--L& R--1& R--2& R--L & R--1 & R--2 & R--L\\ 
        \hline
        \textbf{T5}$_{256+256}$ & 26.49	&9.80	&24.60&	26.94&	10.76&	25.17& 29.00&	10.79&	26.70&	28.53&	10.77&	26.47& 27.74&	10.53	&25.74\\
        \textbf{T5}$_{512}$ & \textbf{28.72}&	\textbf{11.14}&	\textbf{26.59}&	29.03&	12.07&	27.21& 31.39&	12.37&	28.87&	30.96	&12.65	&28.80& 30.03	&12.06	&27.87 \\
        
        \textbf{BART} & 28.40&10.96&	26.35&		28.48&	11.80&	26.72& 31.25&	12.29&	28.79&	30.62&	12.16&	28.51& 29.69	&11.80&	27.59\\
        
        \hline
        \textbf{CodeBERT} & 25.22	&9.31& 	23.62& 	25.17	& 10.28& 	23.85&  28.16& 	10.75	& 26.36	& 27.49& 	10.67& 25.85& 26.51&	10.25&	24.92\\
        \textbf{UnixCoder} & 25.28	&9.34	&23.58&	25.54&	10.49	&24.17 &27.49	&10.00	&26.10	&27.84&	10.66&	26.08& 19.73	&7.65&	18.52  \\
        \textbf{CodeT5} & 28.64 & 11.10&26.53 & \textbf{29.24} & \textbf{12.38} & \textbf{27.31} &\textbf{31.94} & \textbf{12.71} & \textbf{29.40} & \textbf{31.23} & \textbf{12.74} & \textbf{29.16} & \textbf{30.26} &\textbf{12.23}&\textbf{28.10} \\
        \hline
    \end{tabular}      
    \label{tab:result_PLMs}
\end{table*}

\subsection{Implementation Details}\label{sec:implementation}
The implementation of \tool utilizes Transformers,\footnote{\url{https://github.com/huggingface/transformers}} 
based on the checkpoint of the 
CodeT5--base model, which features 12 layers for both encoder and decoder, each with a hidden size of 768.
The input sequences are tokenized using the Byte-Pair Encoding algorithm, characterized by a vocabulary size of 32,100.
Essentially, memory consumption increases quadractically (n$^2$) with respect to the input length. Thus, these sequences are either truncated or padded to a maximum length of 512, so as to optimize the memory. 
The batching is carried out with a size of 4.
We employ the \texttt{AdamW} optimizer, with a default learning rate of $5e^{-5}$ for both the initial fine-tuning and the self--improvement stages. 
The model training lasts 8 epochs during the fine-tuning stage and 4 epochs in the self--improvement phase. We set the number of title candidates to 20 and 30 for the self--improvement and post--ranking processes, respectively.

For other PTMs including BART~\cite{lewis2020bart}, T5~\cite{raffel2020exploring}, CodeBERT~\cite{feng2020codebert} and UnixCoder~\cite{guo2022unixcoder}, the parameters are initialized from \texttt{bart-base}, \texttt{t5-base},  \texttt{codebert-base} and \texttt{unixcoder-base-nine}.
All of the above models can be found on Hugging Face.\footnote{\url{https://huggingface.co/models}} All of our experiments have been carried out on PyTorch 
on a single GPU RTX 2080 Ti with 12GB RAM. The evaluation metrics are computed on experimental results using the \texttt{rouge} library.\footnote{\url{https://github.com/pltrdy/rouge}}

\section{EMPIRICAL RESULTS} 
\label{sec:Results}

\begin{table*}[t!]
    \centering
    \small
    \caption{Effectiveness of \tool compared to state-of-the art baseline models on title generation for \SO posts.}
    \begin{tabular}{l|ccc|ccc|ccc|ccc|ccc}
        \hline
        \multirow{2}{*}{\textbf{Model}} & \multicolumn{3}{c|}{\textbf{Java}} & \multicolumn{3}{c|}{\textbf{C\#}} & \multicolumn{3}{c|}{\textbf{Python}} & \multicolumn{3}{c|}{\textbf{JavaScript}} & \multicolumn{3}{c}{\textbf{Average}}\\
        \cline{2-16} & R--1& R--2& R--L& R--1& R--2& R--L& R--1& R--2& R--L& R--1& R--2& R--L & R--1 & R--2 & R--L\\ 
        \hline
        \textbf{\texttt{Code2Que}}~\cite{gao2021code2que} & 25.25&	9.17&	23.57&	26.55&	10.57&	25.01& 27.93&	10.36&	25.87& 27.50&	10.40&	25.70& 26.81&10.13 & 25.04\\
        \textbf{\texttt{CCBERT}}~\cite{zhang2022improving} & 26.25&	9.65	&24.57&	26.00&	10.76&	24.72 &28.55	&10.69	&26.54	&28.34	&10.83	&26.64& 27.29&	10.48&	25.62\\
        \textbf{\ST}~\cite{liu2022sotitle} & 26.49	&9.80	&24.60&	26.94&	10.76&	25.17& 29.00&	10.79&	26.70&	28.53&	10.77&	26.47& 27.74&	10.53	&25.74\\
        \textbf{\texttt{M3NSCT5}}~\cite{zhang2023diverse} & 27.90 & 10.22 & 25.52 & 28.77 & 11.55 & 26.53 & 31.35 & 11.80 & 28.44 & 30.39 & 11.65 & 27.96 &29.60&	11.31	&27.11\\
        \hline
        \textbf{\GPT} & 25.68&	7.62&	22.52&	25.21&	7.80&	22.69& 28.48&	8.92& 25.06&	27.15&	8.19& 24.11& 26.63& 8.13 & 23.60\\
        \hline
        \textbf{\tool} & \textbf{30.48}	&\textbf{11.52}	&\textbf{27.94}	&\textbf{30.89}	&\textbf{12.81}	&\textbf{28.60}& \textbf{33.64}&	\textbf{13.00}&	\textbf{30.63}&	\textbf{33.14}	&\textbf{13.24}	&\textbf{30.55}&\textbf{ 32.04}&\textbf{ 12.64} &\textbf{ 29.43} \\
        \hline
    \end{tabular}      
    \label{tab:result_baseline}
\end{table*}

This section reports and analyzes the results obtained from the 
experiments 
to answer the research questions 
in Section~\ref{sec:ResearchQuestions}.

\subsection{\rqone}\label{sec:rq1}
Table~\ref{tab:result_PLMs} shows the performance metrics of five different Pre-trained Models (PTMs) across four different programming languages, \ie Java, C\#, Python, and JavaScript. 
For each PTM, we utilized a consistent input modeling approach, as elaborated in Section~\ref{sec:plms}. 
In our research, we handle the tokenized input sequences with a maximum length of 512 for all the PTMs. 
However, our settings slightly differ from the previous configurations suggested by Liu et al.~\cite{liu2022sotitle}.
Instead of combining 256 tokens from the description and code, we adopt a different strategy where we merge the description and code and then truncate the total to 512 tokens, placing greater emphasis on the description than the code. In fact, the number of 512 tokens is defined by the underlying Transformer architecture, and this might have an impact on long posts, \ie those with a lot of code and text.
Considering that Liu et al.’s research~\cite{liu2022sotitle} was mainly concentrated on fine-tuning the T5 model, we also present the outcomes from their study, specifically noting them as $T5_{256+256}$.
The two first rows in Table~\ref{tab:result_PLMs} demonstrate that $T5_{512}$ notably surpasses $T5_{256+256}$, as used by Liu et al.~\cite{liu2022sotitle}, across all programming languages in every performance metric. This implies that the problem description offers more significant information for title generation than the code snippet. We, therefore, apply this settings for all the other PTMs.

As can be seen in Table~\ref{tab:result_PLMs}, the fine-tuned CodeT5 model outperforms all other models in terms of performance metrics for C\#, Python and JavaScript, and it is comparable to the T5 model (labelled as $T5_{512}$ with our settings) for Java.
This superior performance of CodeT5 can be attributed to its architecture, which is based on 
the T5 model but with a specialized emphasis on both natural and programming languages. This allows CodeT5 to more effectively understand the semantics of code snippets. In fact, on average, the CodeT5 model demonstrates enhancements of 0.8\%, 1.9\%, 12.8\%, and 51.7\% in ROUGE-L scores compared to the T5, BART, CodeBERT, and UnixCoder models, respectively.


\find{\textbf{Answer to RQ$_1$:} The fine-tuned CodeT5 model achieves the best performance in comparison to that of the other PTMs for \SO  post title generation.}

\begin{table*}[t!]
    \centering
    \small
    \caption{Ablation Study. 
    $\text{\textbf{\tool}}_{w/o\,PR}$, $\text{\textbf{\tool}}_{w/o\,SI}$ 
    and $\text{\textbf{\tool}}_{w/o\,SI+PR}$ denote three variants of \tool without Post Ranking (PR), Self Improvement (SI) and without both PR and SI, respectively.}
    \begin{tabular}{l|ccc|ccc|ccc|ccc|ccc}
        \hline
        \multirow{2}{*}{\textbf{Model}} & \multicolumn{3}{c|}{\textbf{Java}} & \multicolumn{3}{c|}{\textbf{C\#}} & \multicolumn{3}{c|}{\textbf{Python}} & \multicolumn{3}{c|}{\textbf{JavaScript}} & \multicolumn{3}{c}{\textbf{Average}}\\
        \cline{2-16} & R--1& R--2& R--L& R--1& R--2& R--L& R--1& R--2& R--L& R--1& R--2& R--L & R--1 & R--2 & R--L\\ 
        \hline
        $\text{\textbf{\tool}}_{w/o\,SI+PR}$ & 28.64& 11.10& 26.53& 29.24& 12.38& 27.31& 31.94& 12.71& 29.40& 31.23 & 12.74 & 29.16& 30.26&12.23 &28.10 \\
        
        $\text{\textbf{\tool}}_{w/o\,PR}$ & 28.93& 11.26& 26.89& 29.27& 12.46& 27.45& 32.17& 12.78& 29.64& 31.77 &12.89 &29.59 & 30.54 &12.35&28.39 \\
        $\text{\textbf{\tool}}_{w/o\,SI}$ & 29.71& 11.29& 27.35& 29.86& 12.46& 27.78& 33.05& 12.97& 30.20& 32.34 &12.90 &29.95 & 31.24 &12.41&28.82 \\
         \textbf{\tool} & \textbf{30.48}	&\textbf{11.52}	&\textbf{27.94}	&\textbf{30.89}	&\textbf{12.81}	&\textbf{28.60}&	\textbf{33.64}&	\textbf{13.00}&	\textbf{30.63}&	\textbf{33.14}	&\textbf{13.24}	&\textbf{30.55}&\textbf{ 32.04}&\textbf{ 12.64} &\textbf{ 29.43} \\
        \hline
    \end{tabular}      
    \label{tab:result_ablation}
\end{table*}

\subsection{\rqtwo}~\label{sec:rq2}

\begin{figure}[t]
    \centering
    \includegraphics[width=0.40\textwidth]{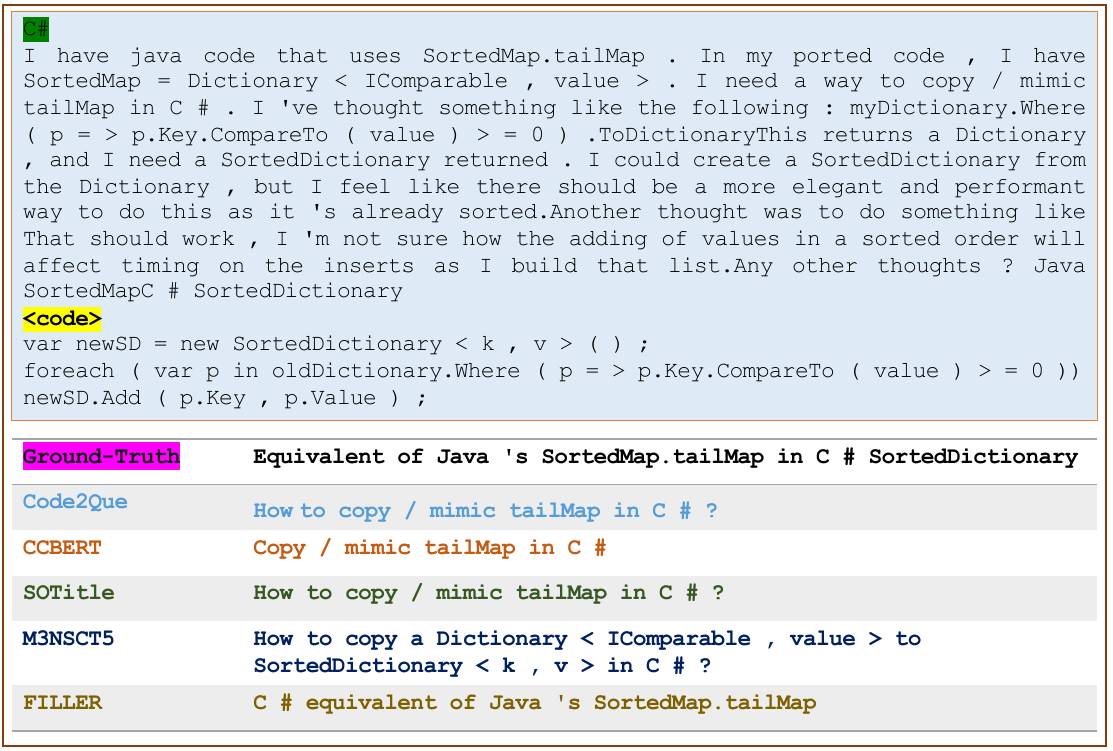}
    \caption{The title generated by \tool compared to those from the baselines for a question post example.}
    \label{fig:rq2-ex} 
    \vspace{-.2cm}
\end{figure}

To assess the effectiveness of our proposed approach in generating \SO post titles, we compare \tool with four recent and well-established baselines, namely \texttt{Code2Que}~\cite{gao2021code2que}, \texttt{CCBERT}~\cite{zhang2022improving}, \texttt{SOTitle}~\cite{liu2022sotitle} and \texttt{M3NSCT5}~\cite{zhang2023diverse}.
The comparison results across four programming languages are reported in Table~\ref{tab:result_baseline}.
In general, \tool significantly outperforms the other baselines in terms of all performance measures for all programming languages. Particularly, the performance comparison reveals that text-to-text baseline models such as \texttt{SOTitle}, \texttt{M3NSCT5}, and \tool outperform \texttt{Code2Que} and \texttt{CCBERT} models. Notably, \texttt{Code2Que}, relying on a \texttt{sequence-to-sequence} architecture, proves to be the least effective for title generation. For instance, in terms of ROUGE–1, the average improvement across all programming languages, when compared to \texttt{Code2Que}, is 1.8\% for \texttt{CCBERT}, 3.5\% for \texttt{SOTitle}, 10.4\% for \texttt{M3NSCT5}, and 19.5\% for \tool.
The \texttt{SOTitle} model exhibits slightly better performance than \texttt{CCBERT} but is less effective when compared to \texttt{M3NSCT5} and \tool.
This observation is supported by the fact that \ST utilizes fine-tuned T5, which outperforms CodeBERT--the backbone model of \texttt{CCBERT}, in generating 
post titles, as indicated in the findings of the previous question.

Figure~\ref{fig:rq2-ex} shows the titles generated by \tool and the baseline models for a question post example related to the programming language C\#.\footnote{The post is available online in this link: \url{https://bit.ly/3Hn3o2Z}} 
In this example, the titles produced by \texttt{Code2Que}, \texttt{CCBERT}, and \ST successfully incorporate important keywords such as ``copy/mimic'', ``tailMap'' and ``C\#''. However, these titles lack the nuanced details of the underlying problem. 
In contrast, \texttt{M3NSCT5} and \tool offer a more precise expression of key information in the post. 
The title generated by \tool closely aligns with the semantic meaning of the ground-truth title. 
While \texttt{M3NSCT5} aims to enhance diversity by expanding the exploration space during token generation, often resulting in longer sentences, \tool focuses on increasing the likelihood of producing high-quality titles by exploring a wide range of candidates and evaluating their relevance to the post. 
Indeed, our proposed approach outperforms \texttt{M3NSCT5} across all performance metrics, achieving an average improvement of 8.2\% on ROUGE--1, 11.8\% on ROUGE--2, and 8.6\% on ROUGE--L.


By evaluating \tool's performance against ChatGPT with \GPT, as 
in Table~\ref{tab:result_baseline}, we see that \tool significantly outperforms \GPT in all the metrics, including ROUGE--1, ROUGE--2, and ROUGE--L, in various 
languages. Specifically, the average improvements are 20.3\% for ROUGE--1, 55.5\% for ROUGE--2, and 24.7\% for ROUGE--L.

\find{\textbf{Answer to RQ$_2$:} \tool consistently outperforms the four state-of-the-art baseline models, including \texttt{SOTitle}, \texttt{Code2Que}, \texttt{CCBERT}, \texttt{M3NSCT5} and \GPT in generating tiles for \SO posts consisting of code written in four different programming languages.}

\subsection{\rqthree}~\label{sec:rq3}


In this investigation, we conduct an ablation study to dissect the factors contributing to the improvement in performance. 
We examine three specific scenarios: (1) Excluding only the Post-Ranking component, denoted as $\text{\tool}_{w/o\,PR}$; (2) 
Leaving out the Self-Improvement component, denoted as $\text{\tool}_{w/o\,SI}$ ; and (3) 
Removing both the Self-Improvement and Post-Ranking components, denoted as  $\text{\tool}_{w/o\,SI+PR}$.
Table~\ref{tab:result_ablation} shows the obtained results from these experiments.

It can be observed that, without the presence of both Self-Improvement and Post-Ranking, \tool experiences a consistent decline across all the 
metrics for various 
languages. For instance, in terms of ROUGE-1, $\text{\tool}_{w/o\,SI+PR}$ exhibits a decrease of 6.4\% for Java, 5.6\% for C\#, 5.3\% for Python, and 6.1\% for JavaScript, leading to an average reduction of 5.9\%.
The results are also consistent for ROUGE-2 and ROUGE-L.
In the second scenario, excluding the Post-Ranking component while retaining Self-Improvement slightly boosts performance compared to the model lacking both components. Specifically, with Self Improvement active, there is an average increase of 0.93\% for ROUGE-1, 0.98\% for ROUGE-2, and 1.03\% for ROUGE-L, compared to $\text{\tool}{w/o,SI+PR}$. In the third scenario, including Post-Ranking but omitting Self-Improvement leads to even better performance than the second scenario. Specifically, $\text{\tool}{w/o,SI}$ shows an average improvement of 2.3\% for ROUGE-1, 0.49\% for ROUGE-2, and 1.5\% for ROUGE-L compared to $\text{\tool}_{w/o,PR}$, highlighting the value of generating and ranking diverse titles to improve the model's output quality.
In summary, our experiments indicate that the exclusion of either component from \tool results in a decline in performance when generating 
post titles. This underscores the significance of both Self-Improvement and Post-Ranking for the effective operation of \tool.


\begin{table}[h!]
	\centering
	\small
	\caption{Analysis of different $\mathbb{K}$-values in post ranking.}
	\begin{tabular}{l|ccc}
	\hline
	\multirow{2}{*}{\textbf{Top-k Ranking}} & 
	\multicolumn{3}{c}{\textbf{Average}}\\
	\cline{2-4} & 
	R--1 & R--2 & R--L\\ 
	\hline
	\textbf{$\mathbb{K}=10$} 
	&31.46	&12.45 &29.02
	\\
	\textbf{$\mathbb{K}=20$} 
	&31.82	&12.55	&29.23
	\\
	\textbf{$\mathbb{K}=30$} 
	&32.04	&\textbf{12.64}	&29.43
	\ \\
	\textbf{$\mathbb{K}=40$} 
	&32.06	&12.62	&29.44
	\\
	\textbf{$\mathbb{K}=50$} 
	&\textbf{32.13}	&12.56	&\textbf{29.45}
	\\
	\hline
\end{tabular}      
\label{tab:k-comparison}
\end{table}

\begin{table*}[t!]
	\footnotesize
	\centering
	\caption{Examples of the evaluation scores from the user study.}
        \vspace{-.2cm}
	\label{tab:RecommendationExamples}
	\begin{tabular}{|c|p{3.0cm}|p{3.0cm}|p{0.24cm}|p{0.24cm}|p{3.0cm}|p{0.24cm}|p{0.24cm}|p{3.0cm}|p{0.24cm}|p{0.24cm}|} \hline
	     &  & \multicolumn{3}{c|}{\textbf{\ST}}  &  \multicolumn{3}{c|}{\textbf{\GPT}} & \multicolumn{3}{c|}{\textbf{\tool}} \\        \hline 
	 No. & \textbf{Ground-truth title} & \textbf{Generated title} & R$_1$ & R$_2$ & \textbf{Generated title} & R$_1$ & R$_2$ & \textbf{Generated title} &  R$_1$ & R$_2$ \\        \hline
	 
        1 & React-highcharts scrollbar feature not working   & react-highchart scrollbar not working	& 2 & 4  & How to handle errors and rewrite mapping logic using Ramda in functional programming for indexing target columns in CSV header? & 4 & 1 & react-highcharts bar scrollbar not working & 2 & 4 \\ \hline
        
        2 & Why is 'this' undefined in the debugger but printable in a console.log?   & Why does console.log set a breakpoint on console.log?	& 3 & 1  & What is the purpose of the "(0, foo)()" syntax in the JavaScript code snippet provided? & 3 & 1 & Why is `this` undefined in console.log?  & 4 & 3 \\ \hline
        
        3 &  Why does an unhandled exception in this background thread not terminate my process?  & 	  Unhandled exception in foreground and background threads & 4 & 2  & How to achieve constant rotation of a character's head without flipping the character when rotated more than 180 degrees in Unity? & 3 & 1 & Why does an unhandled exception in a background thread not cause the process to terminate ? & 3 & 4  \\ \hline
        
        4 &  "list.toArray (T [ ] a), what if the ``T'' is a ``runtime type''?"  & How to convert a list into an array if it is a  runtime type''?	& 3 & 2  & Java 9 issue with casting toArray in Arrays.asList causing code to not work & 3 & 1 & List.toArray (T [ ] a), what if the T is a ``runtime type''? & 2 & 4 \\ \hline
        
	\end{tabular}	
\end{table*}

For a more in-depth analysis of the effectiveness of Post-Ranking, 
we delve into the impact of selecting the $\mathbb{K}$ value during the generation of title candidates. Table~\ref{tab:k-comparison} presents a comparative analysis of various $\mathbb{K}$-values, ranging from $10$ to $50$, for clarity. A list of $50$ items is quite lengthy, thus representing the upper limit worth considering. The empirical data shows consistent improvements in ROUGE-1 and ROUGE-L scores across all 
languages when the $\mathbb{K}$-value is increased from 10 to 50.
This aligns with our expectations due to the maximization-based objective of the beam search algorithm during the decoding stage, as well as the intricacies of the TextRank algorithm in the post-ranking process. 
The beam search algorithm prioritizes tokens with the highest probability, emphasizing key tokens from the input. Consequently, increasing the number of sample candidates enhances the ability to capture these key tokens. On the other hand, TextRank tends to assign high ranking scores to titles containing common key tokens from others, leading to the selection of longer titles with a greater number of important tokens.
However, longer titles may have a detrimental effect on ROUGE-2. Table~\ref{tab:k-comparison} shows that compared to the $\mathbb{K}$-value of 10, the $\mathbb{K}$-value of 50 yields an average improvement of 1.65\% for ROUGE-1 and 1.07\% for ROUGE-L. 
Conversely, the $\mathbb{K}$-value exceeding 30 results in a performance decline in terms of ROUGE-2. 
Balancing between speed and quality is crucial, making the $\mathbb{K}$-value of 30 the preferred choice, considering the impracticality of generating an excessive number of sample titles.

\vspace{.2cm}
\find{
\textbf{Answer to RQ$_3$:} Without Self-Improvement and Post-Ranking, the performance of \tool decreases by 5.9\% and 3.4\% and 4.7\% in terms of ROUGE-1, ROUGE-2 and ROUGE-L, respectively. A $\mathbb{K}$ value set to 30 for the Post-Ranking process provides a balanced compromise between the speed and quality of the generated titles. 
All the components of \tool contribute positively to its performance.}

\subsection{\rqfour}~\label{sec:rq4}

Before analyzing the results, we extract some evaluation examples from the user study, and show them in Table~\ref{tab:RecommendationExamples} to illustrate how the outcome looks like with respect to \emph{Readability} (R$_1$) and \emph{Relevance} (R$_2$).\footnote{Due to space limit, we present only 4 representative examples. Interested readers are kindly referred to the replication package \cite{FILLERartifacts} for more details.} It is evident that \GPT usually generates \emph{verbal} titles, containing a lot of text (R$_1$ is considerably high) but not relevant to the ground-truth ones, \ie all the samples get 1 as R$_2$. \ST is better compared to \GPT in recommending highly relevant titles. Among others, \tool is able to produce more \emph{factual} titles, being close to the original ones, \ie having 3 and 4 as the Relevance scores. This is also the case by the remaining samples, as we show in the following analysis.

\begin{figure}[h!]
	\centering
	\includegraphics[width=0.45\textwidth]{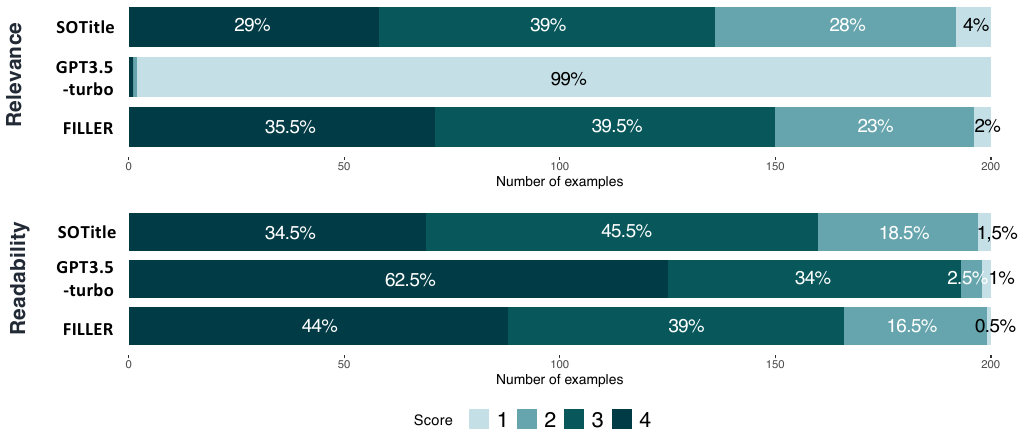}
	\caption{Evaluation results from the user study.} 
	\vspace{-.2cm}
	\label{fig:rq4-human-eval} 
\end{figure}

Figure~\ref{fig:rq4-human-eval} shows the evaluation results on human study of \ST, \GPT, and \tool.
In terms of Relevance, the titles generated by \GPT significantly deviate from the content of the 
post and the corresponding ground-truth titles, with 99\% of the examples receiving a score of $1$ for the predicted titles. Essentially, \tool performs better, achieving the top score of 4 in 35.5\% of cases, which is 6.5\% higher than \ST, showcasing that our model more precisely encapsulates the content of SO posts. Both \texttt{SOTitle} and \tool effectively capture the essential points of the posts, with relevance scores of 3 and 2 in 67\% and 62.5\% of samples, respectively.
Regarding Readability, \GPT demonstrates its strength with 62.5\% of examples scoring the highest mark of 4, highlighting its ability to craft titles that are well-written and appealing to human readers.
The titles from both \tool and \ST are clear and easy to comprehend, with readability scores of 4 and 3 in 83\% and 80\% of the samples, respectively.

Finally, we validate the null hypothesis that there are no statistical differences between the performance of \tool and that of \ST as well as of \GPT. Computing the Wilcoxon rank sum test \cite{Wilcoxon1992} on the scores obtained by each pair of tools, \ie \tool vs. \ST and \tool vs. \GPT, we see that 
p-values for R$_1$ and R$_2$ are smaller than 1.06e-05 and 2.2e-16, respectively. 
Considering the 95\% significance level (p-value < 0.05), it is evident that the p-values are lower than 5e-02. Thus, we reject the null hypothesis and conclude that the performance improvement obtained by \tool is statistically significant.



\vspace{.2cm}
\find{\textbf{Answer to RQ$_4$:} The user study demonstrates that \tool is capable of generating high-quality titles, which capture well the intrinsic characters of \SO posts, resembling those given by humans. The obtained performance gain is statistically significant.}

\subsection{Threats to validity}

There might be the following threats to validity of our findings.

\smallskip
\noindent
\textbf{Internal validity.} This is related to the fact that our evaluation resembles real-world scenarios. We used existing datasets that have been widely used for the same purpose. The comparison with the baselines was done using their original implementation, \ie we ran the tools provided in their replication package, attempting to mitigate any possible threats to internal validity. The user study was conducted following existing guidelines~\cite{DBLP:journals/ese/RoccoRSNR21}, first by anonymizing the posts, and then discussing among the evaluators to aim for sound evaluation, as well as to avoid any bias. A probable threat is the amount of information discarded during the text encoding phase. We truncated the input sentences to 512 tokens, as this was pre-defined by the Transformers library. This can be subject to loss of information if the posts are longer than this threshold.

\smallskip
\noindent       
\textbf{External validity.} This concerns the generalizability of the findings outside this study. In the evaluation, we experimented with datasets covering four different programming languages, \ie Java, C\#, Python, and JavaScript, and our findings are valid for posts containing code in these languages. 

\smallskip
\noindent    
\textbf{Construct validity.} This threat is about the setup and measurement in the study, \ie if they resemble real-world situations. We  attempted to simulate a real scenario of generating titles for \SO posts,  by means of cross validation. In particular, the dataset was split into two independent parts, including a training set and a testing set. This simulates the scenario where the items in the training data correspond to the posts available for the recommendation process. 
    Meanwhile, an item in the testing data corresponds to the post under consideration, and the models are expected to generate a title for each post. 

\section{CONCLUSION AND FUTURE WORK}
\label{sec:Conclusion}

This paper presented an approach named \tool to \SO post title generation. Our tool leverages fine-tuning language model with self-improvement and post-ranking. By incorporating the model's own predictions into the training process, it can learn from its errors, thus mitigating the effects of exposure bias. Moreover, a post-ranking method was employed to produce a variety of sample candidates, subsequently selecting the most suitable one, thus increasing the likelihood of generating more relevant titles. An empirical evaluation using real-world datasets with posts covering four different programming languages shows that \tool obtains a promising recommendation performance, thus outperforming four state-of-the-art baselines for post title recommendation. 

For future work, we plan to test \tool with data from different sources and posts related to other programming languages. Moreover, we will also incorporate various post ranking mechanisms to improve the recommendation performance further.

\begin{acks} 
	
	This work has been partially supported by the EMELIOT national research project, which has been funded by the MUR under the PRIN 2020 program (Contract 2020W3A5FY). 
	The work has been also partially supported by the European Union--NextGenerationEU through the Italian Ministry of University and Research, Projects PRIN 2022 PNRR \emph{``FRINGE: context-aware FaiRness engineerING in complex software systEms''} grant n. P2022553SL. We acknowledge the Italian ``PRIN 2022'' project TRex-SE: \emph{``Trustworthy Recommenders for Software Engineers,''} grant n. 2022LKJWHC.
\end{acks}

\bibliographystyle{ACM-Reference-Format}
\bibliography{main}


\end{document}